\documentclass[showpacs,amsmath,12pt,prd,onecolumn,nofootinbib]{revtex4}
\usepackage{graphicx} 
\usepackage{amsmath}
\usepackage{amssymb}
\usepackage{color}
\usepackage{tikz}
\usepackage{pgfplots}
\usepackage{tabularx} 
\usepackage{tikz}
\usetikzlibrary{arrows.meta,positioning,shadows}
\begin{document}
\title{A Modified Gravitational Theory of the Matter Sector of the Type $\phi(R,T)\mathcal{L}_{m}$}

\author{Ginés R. Pérez Teruel}
\email{gines.landau@gmail.com}
\affiliation{Consellería de Educación, Cultura, Universidades y Empleo, Ministerio de Educación y Formación Profesional, Spain}

\author{Antonio Peña Peña}
\email{antoniopenapena52@gmail.com}
\affiliation{Higher Polytechnic School, Autonomous University of Madrid, 28049-Cantoblanco, Spain}
\begin{abstract}
We investigate a modified gravity framework where the geometric Einstein--Hilbert sector remains untouched while the matter Lagrangian is weighted by a nontrivial function $\phi(T)$ of the energy--momentum trace. 
Unlike $f(R,T)$ or $f(R,\mathcal L_m)$ theories, this construction alters how matter curves spacetime without introducing extra geometric degrees of freedom, thereby remaining consistent with local tests of gravity. 
Physically, the factor $\phi(T)\mathcal L_m$ can be interpreted as an effective renormalization of the matter sector, relevant at high densities and smoothly reducing to GR at low densities. 
Within this setup we identify a robust window in parameter space leading to a smooth and nonsingular cosmological bounce, with bounded density, finite $\dot H>0$ at the bounce, and preservation of the infrared limit where $\Lambda$CDM is recovered. 
This mechanism provides a natural route to singularity resolution while retaining the empirical successes of standard cosmology.
\end{abstract}

\maketitle

\section{Introduction and Motivation}
General Relativity (GR) remains the standard paradigm for describing the gravitational interaction, but it is well known that it cannot provide the final word on the problem of gravity. The theory inevitably predicts spacetime singularities, a feature that signals its limitations and points to the need for new physics at high energies \cite{Hawking1970,Penrose1965,Wald1984}. It is therefore widely believed that GR is incomplete and must be generalized by a more fundamental framework that remains valid in regimes where GR breaks down. Several avenues have been proposed, ranging from inflationary cosmology \cite{Starobinsky1980,Guth1981,Linde1983} and scalar--tensor theories such as Horndeski gravity \cite{Horndeski1974,DeFelice2010,Langlois2019}, to quantum gravity inspired models including loop quantum cosmology \cite{Ashtekar2006,Bojowald2005} and string-motivated bouncing scenarios \cite{Novello2008,Brandenberger2017,Battefeld2015,Cai2012,Cai2014,Brandenberger2016}.  

Over the past decades a wide spectrum of modified gravity frameworks has been developed in order to address the shortcomings of GR. Among the most influential are $f(R)$ theories and their scalar--tensor generalizations \cite{Capozziello2007,Nojiri2011}, as well as extensions such as $f(Q)$ gravity, built on the non-metricity scalar $Q$ \cite{Jimenez2018,Heisenberg2019}, and Palatini-inspired approaches like $f(R,Q)$, which can avoid singularities while keeping the field equations of second order \cite{Olmo2011,OlmoRubiera2015,Olmo2016}. At the phenomenological level, modified matter couplings of the $f(R,T)$ and $f(R,\mathcal L_m)$ type have been extensively explored both in cosmology \cite{Harko2011,Moraes2016,Shabani2013,Odintsov2013,Bertolami2007,Bertolami2008,Harko2010,BarrosoVarelaBertolami2024,BarrosoVarelaBertolami2025} and in astrophysics, where they provide viable models of compact stars and anisotropic fluids \cite{Rahaman2012,Rahaman2014,Singh2014,Singh2019,Singh2020,Singh2021}. This ``zoo'' of modified gravity models illustrates the diversity of possible departures from GR, but also highlights the importance of identifying frameworks that are both phenomenologically robust and free from severe stability issues.  

An important motivation for our approach is to clarify how it differs from the better-known frameworks of $f(R,T)$ and $f(R,\mathcal L_m)$ theories. In those extensions, the modification acts primarily on the geometric sector of the action, leading to additional curvature-dependent terms in the field equations that explicitly involve either the energy--momentum tensor or the matter Lagrangian. In contrast, in the present model the geometric sector remains strictly Einsteinian, and the modification is introduced exclusively as a nontrivial weight $\phi(R,T)$ multiplying the matter Lagrangian. This structural change has several important consequences. Since the purely geometric part of the action is untouched, no extra propagating spin-0 or spin-2 modes arise in vacuum, which automatically avoids the strongest constraints from local gravity tests that typically affect $f(R)$ and $f(R,T)$ models. Moreover, the mechanism at work is fundamentally different from that of $f(R,\mathcal L_m)$ theories. In the latter, the nonconservation of the energy--momentum tensor originates from explicit curvature--matter couplings of the form $f(R,\mathcal L_m)$. In our case, nonconservation arises instead from the dependence of the prefactor $\phi(T)$ on the trace of the energy--momentum tensor. The resulting energy exchange between matter and geometry is conceptually distinct: the background curvature continues to be governed by pure Einstein--Hilbert dynamics, but the way matter sources curvature is renormalized by the function $\phi(T)$.  

It is also important to address a criticism raised in the literature, namely that modifications which act solely on the matter sector should not be regarded as genuine theories of modified gravity, since they could be reinterpreted within special relativity \cite{Lind,Visser}. Our framework avoids this objection because the weighting $\phi(T)\mathcal L_m$ enters directly into the action integral, thereby modifying the variational principle and the coupling of matter to spacetime curvature. This is not a mere redefinition of matter fields in a fixed background, but a genuine modification of how matter curves spacetime.  

From a physical point of view, the factor $\phi(T)$ multiplying the matter Lagrangian can be interpreted as an effective renormalization of the matter sector in regimes of high density. At low densities one recovers $\phi(T)\to 1$, reproducing standard general relativity. At high densities, however, the weighting introduces new effective interactions that mimic the kind of operators expected in a quantum effective field theory \cite{BirrellDavies,ParkerToms,Donoghue,BOS,BarvinskyVilkovisky}. From this perspective, the model may be viewed as a low-energy effective description of possible high-energy corrections to general relativity, in which the matter sector ``feels'' quantum backreaction through a density-dependent renormalization.  
\section{General Field Equations}
The total action of the theory is given by
\begin{equation}
\displaystyle S=\int\left[\frac{1}{2\kappa}R+\phi(R,T)\mathcal{L}_{m}\right]\sqrt{-g}\,\mathrm {d}^{4}x.
\end{equation}

Where $R= R^{\mu\nu}g_{\mu\nu}$ is the curvature scalar, $\mathcal{L}_{m}$ is the matter Lagrangian, $\kappa=8\pi G$ is Einstein's gravitational constant, $T=T^{\mu\nu}g_{\mu\nu}$ represents the trace of the energy momentum tensor, while $\phi(R,T)$ is a function that reduces to $1$ for GR.  
Variation of the action given by (1) leads to the field equations
\begin{equation}\label{fieldEquations}
R_{\mu\nu}-\frac{1}{2}Rg_{\mu\nu}=\kappa T^{eff}_{\mu\nu}.
\end{equation}    

Where $T^{eff}_{\mu\nu}$ is an effective stress-energy tensor given by
\begin{equation}
T^{eff}_{\mu\nu}=\phi T_{\mu\nu}-2\phi_{R}R_{\mu\nu}\mathcal{L}_{m}-2\Big(g_{\mu\nu}\Box- \nabla_{\mu}\nabla_{\nu}\Big)(\phi_{R}\mathcal{L}_{m})-2\phi_{T}\frac{\delta T}{\delta g^{\mu\nu}}\mathcal{L}_{m}.
\end{equation}
Where, as usual
\begin{equation}\label{StressEnergyTensorDef}
T_{\mu\nu}=-\frac{2}{\sqrt{-g}}\frac{\delta(\sqrt{-g}\mathcal{L}_{m})}{\delta g^{\mu\nu}},
\end{equation}
and we have denoted $\phi_{R}=\partial \phi/\partial R$ and $\phi_{T}=\partial\phi/\partial T$, respectively.\\

Taking the covariant derivative of (3) we have
\begin{equation}
\nabla^{\mu}T^{eff}_{\mu\nu}=0.
\end{equation}
 This equation describes the total energy transfer between matter and geometry. Notice that since $T^{eff}$ is conserved, total energy exchange between matter and geometry is conserved, but the individual sectors do not. Let us define now
 \begin{equation}
 \frac{\delta T}{\delta g^{\mu\nu}}= \frac{\delta (g^{\alpha\beta}T_{\alpha\beta})}{\delta g^{\mu\nu}}=T_{\mu\nu}+\Theta_{\mu\nu}, 
 \end{equation}
 where 
 \begin{equation}
 \Theta_{\mu\nu}=g^{\alpha\beta}\frac{\delta T_{\alpha\beta}}{\delta g^{\mu\nu}}.
 \end{equation}
 Or, in terms of the matter Lagrangian
 \begin{equation}
 \Theta_{\mu\nu}=-2T_{\mu\nu}+g_{\mu\nu}\mathcal{L}_{m}-2g^{\alpha\beta}\frac{\partial^{2}\mathcal{L}_{m}}{\partial g^{\mu\nu}\partial g^{\alpha\beta}}.
 \end{equation}
The field equations will be free of higher order derivatives of the metric coefficients if we require that $\phi(R,T)$ be linear function of $R$, namely $\phi(R,T)=1+\alpha R+\beta g(T)$. Therefore, $\phi_{R}=\alpha$, and the effective stress-energy tensor will read

\begin{equation}
T^{eff}_{\mu\nu}=\phi T_{\mu\nu}-2\alpha R_{\mu\nu}\mathcal{L}_{m}-2\alpha\Big(g_{\mu\nu}\Box- \nabla_{\mu}\nabla_{\nu}\Big)\mathcal{L}_{m}-2\phi_{T}\frac{\delta T}{\delta g^{\mu\nu}}\mathcal{L}_{m}.
\end{equation}

Some comments are in order. In principle, the cosmological constant $\Lambda$ should be included in the action alongside the curvature scalar term. However, its energy scale is negligible compared to the effective corrections introduced by $\phi(R,T)\mathcal L_m$ in the high-density regime. The bounce dynamics analyzed later in this work are therefore insensitive to $\Lambda$. At low densities, on the other hand, the model smoothly reduces to GR with $\Lambda$, recovering the $\Lambda$CDM phenomenology in the infrared.
\paragraph{On the choice of the matter Lagrangian.}
In theories with nonminimal matter couplings, the choice of the matter
Lagrangian $\mathcal L_m$ is known to be nonunique. In the case of perfect
fluids, both $\mathcal L_m = p$ and $\mathcal L_m = -\rho$ are commonly used,
and they are equivalent in minimally coupled general relativity. However,
in the present framework, where $\mathcal L_m$ appears explicitly in the
action through the factor $\phi(T)$, different choices may lead to
quantitative differences in the effective energy--momentum tensor.

In this work we adopt the standard choice $\mathcal L_m = p$, which is widely
used in the literature. We have verified that the qualitative features of the
cosmological dynamics, in particular the existence of a nonsingular bounce
driven by the density dependence of $\phi(T)$, do not rely on this specific
choice. Alternative prescriptions such as $\mathcal L_m = -\rho$ modify the
detailed form of the effective sources but preserve the key mechanism
responsible for the bounce discussed in the next sections.

In particular, even in the dust limit $p=0$, the trace remains $T=-\rho\neq 0$,
so that $\phi(T)$ still induces nontrivial modifications.
\subsection{Connection with other gravities}
\begin{itemize}

\item{\textbf{Case I. $\phi=\phi(T)$. Matter-matter coupling $\phi(T)\mathcal{L}_{m}$.}}\\
When the matter sector of the action is purely weighted by the proper matter itself, that is, $\phi=\phi(T)$, the theory is described by the following field equations
\begin{equation}
R_{\mu\nu}-\frac{1}{2}Rg_{\mu\nu}=\kappa\phi(T)T_{\mu\nu}-2\kappa\phi_{T}(T_{\mu\nu}+\Theta_{\mu\nu})\mathcal{L}_{m}.
\end{equation}
In a next section, we will build an example of a nonsingular, stable cosmological model for this particular case. If the matter sources are characterized by a presureless dust, ($p=0$, $\mathcal{L}_{m}=p=0$), a direct link with other gravity is naturally established. Under these assumptions, a considerable simplification of the field equations is achieved, which read
\begin{equation}
R_{\mu\nu}-\frac{1}{2}Rg_{\mu\nu}=\kappa \phi(T)T_{\mu\nu}.
\end{equation}
This agrees with the field equations of the so-called $\kappa(R,T)$ theory of gravity \cite{Teruel}, with the identification $\kappa(T)=\kappa\phi(T)$.
It is worth emphasizing that the present $\phi(R,T)\mathcal L_m$ framework can be regarded as a natural evolution of the $\kappa(R,T)$ gravity model. This latter formulation is intrinsically non--Lagrangian, as the modification acts directly at the level of the field equations by rescaling $T_{\mu\nu}$. 
In contrast, the present $\phi(R,T)\mathcal L_m$ theory admits a variational formulation in which the matter Lagrangian is weighted by a nontrivial function $\phi(R,T)$, leading to a well-defined effective energy--momentum tensor. 
From this perspective, $\phi(R,T)\mathcal L_m$ provides a more fundamental realization of the same conceptual idea: to introduce high-energy corrections by renormalizing the matter coupling to gravity, while leaving the pure Einsteinian geometry intact.

\item {\textbf{Case II. $\phi=\phi(R)$. Curvature-matter coupling: $\phi(R)\mathcal L_{m}.$}}\\
The field equations for this special case reduce to
\begin{equation}
R_{\mu\nu}-\frac{1}{2}Rg_{\mu\nu}=\kappa\phi(R) T_{\mu\nu}-2\phi_{R}R_{\mu\nu}\mathcal{L}_{m}-2\Big(g_{\mu\nu}\Box- \nabla_{\mu}\nabla_{\nu}\Big)(\phi_{R}\mathcal{L}_{m}).
\end{equation}
This particular case agrees with the theory already studied by Bertolami et al. \cite{Bert}. No further comments will be made, referring the readers to the aforementioned work.
\end{itemize}

\subsection{Extra force and weak-field behavior}

In the present framework, we restrict to the case $\phi=\phi(T)$.
The nonminimal coupling then implies a nonvanishing divergence of the
matter energy--momentum tensor, $\nabla^\mu T_{\mu\nu} \neq 0$, which
can be interpreted as inducing deviations from geodesic motion of the
fluid flow.

To make this explicit, we consider a perfect fluid with
\begin{equation}
T_{\mu\nu}=(\rho+p)u_\mu u_\nu + p\,g_{\mu\nu},
\qquad \mathcal L_m = p,
\end{equation}
for which
\begin{equation}
\Theta_{\mu\nu}=-2T_{\mu\nu}+p\,g_{\mu\nu},
\end{equation}
and therefore
\begin{equation}
T_{\mu\nu}+\Theta_{\mu\nu}=-(\rho+p)\,u_\mu u_\nu.
\end{equation}

Starting from the modified conservation equation and projecting orthogonally
to the four-velocity using $h^\alpha{}_\nu=\delta^\alpha{}_\nu+u^\alpha u_\nu$,
one obtains the modified Euler equation
\begin{equation}
a^\alpha \equiv u^\mu\nabla_\mu u^\alpha
=
-\frac{\phi\,h^{\alpha\nu}\nabla_\nu p
+p\,h^{\alpha\nu}\nabla_\nu\phi}
{(\rho+p)\,(\phi+2p\,\phi_T)},
\end{equation}
valid whenever $\phi+2p\,\phi_T\neq0$.
This reduces to the standard GR expression
\begin{equation}
a^\alpha_{\rm GR}=-\frac{h^{\alpha\nu}\nabla_\nu p}{\rho+p}
\end{equation}
in the limit $\phi\to1$ and $\phi_T\to0$.

The deviation from GR can be interpreted as an effective extra force,
defined by $a^\alpha = a^\alpha_{\rm GR} + f^\alpha_{\rm extra}$. A direct
comparison yields
\begin{equation}
f^\alpha_{\rm extra}
=
\frac{p\,\phi_T}{(\rho+p)(\phi+2p\phi_T)}
\left(2\,h^{\alpha\nu}\nabla_\nu p - h^{\alpha\nu}\nabla_\nu T\right).
\end{equation}

For a barotropic fluid $p=w\rho$ with constant $w$, this simplifies to
\begin{equation}
f^\alpha_{\rm extra}
=
\frac{w(1-w)\,\phi_T}{(1+w)(\phi+2w\rho\,\phi_T)}
\,h^{\alpha\nu}\nabla_\nu\rho.
\end{equation}

\paragraph{Low-density regime.}
To make contact with physical scales, we now specialize to the rational ansatz
\begin{equation}
\phi(T)=1+\frac{\beta T}{1+\gamma T},
\end{equation}
for which
\begin{equation}
\phi_T(T)=\frac{\beta}{(1+\gamma T)^2}.
\end{equation}
In the weak-coupling, low-density regime defined by
\begin{equation}
|\beta T|\ll1,
\qquad
|\gamma T|\ll1,
\end{equation}
one has
\begin{equation}
\phi\simeq1,
\qquad
\phi_T\simeq\beta,
\end{equation}
so that the extra force reduces to
\begin{equation}
f^\alpha_{\rm extra}
\simeq
\frac{w(1-w)\beta}{1+w}\,h^{\alpha\nu}\nabla_\nu\rho.
\end{equation}

For $w\neq0$, comparison with the GR Euler acceleration,
\begin{equation}
a^\alpha_{\rm GR}=-\frac{w}{(1+w)\rho}\,h^{\alpha\nu}\nabla_\nu\rho,
\end{equation}
yields the dimensionless estimate
\begin{equation}
\frac{|f_{\rm extra}|}{|a_{\rm GR}|}
\simeq
(1-w)\,|\beta|\,\rho.
\end{equation}

This identifies the scale
\begin{equation}
\rho_\star \sim |\beta|^{-1},
\end{equation}
which controls the magnitude of the extra force in the low-density regime.
Hence, whenever $\rho\ll \rho_\star$, deviations from geodesic motion are
strongly suppressed. For the barotropic relation $T=\tau\rho$ with
$\tau=-1+3w$, the onset of the nonlinear regime of the coupling is instead
controlled by the trace scales $|\beta\tau\rho|\sim1$ and $|\gamma\tau\rho|\sim1$.

In this sense, the extra force is naturally negligible in weak-field,
low-density environments, while the modification becomes relevant only in
the high-density regime where the bounce occurs. In the dust limit
$p=0$, one has $f^\alpha_{\rm extra}=0$, so that pressureless matter
follows geodesics even though the gravitational coupling remains modified
through the trace dependence of $\phi(T)$.

\section{Comparison with other frameworks}
\label{sec:comparison}

It is useful to place the present proposal in the broader landscape of modified matter--geometry couplings, in particular relative to the two best studied alternatives, namely $f(R,T)$ and $f(R,\mathcal L_m)$. Both share with our framework the feature of violating the standard conservation law $\nabla^\mu T_{\mu\nu}=0$, but the origin and severity of this violation differ in essential ways.  

In $f(R,T)$ gravity, introduced in \cite{Harko2011}, the Einstein--Hilbert action is replaced by a function $f(R,T)$ depending on both the scalar curvature and the matter trace. The metric equations then contain the derivatives $f_R=\partial f/\partial R$ and $f_T=\partial f/\partial T$, which introduce higher-order terms on the geometric side. The Bianchi identity implies a modified conservation law of the form
\begin{equation}
\nabla^\mu T_{\mu\nu}
=\frac{f_T}{\kappa-f_T}\left[
\big(T_{\mu\nu}+\Theta_{\mu\nu}\big)\nabla^\mu\!\ln f_T
+\nabla^\mu\Theta_{\mu\nu}-\tfrac{1}{2}g_{\mu\nu}\nabla^\mu T
\right],
\end{equation}
where $\Theta_{\mu\nu}\equiv g^{\alpha\beta}\delta T_{\alpha\beta}/\delta g^{\mu\nu}$. In general one has
\begin{equation}
\nabla_\mu f_T = f_{TT}\,\nabla_\mu T + f_{RT}\,\nabla_\mu R,
\end{equation}
so curvature gradients explicitly enter the nonconservation law only in the non--separable case $f_{RT}\neq 0$. For separable models, $f(R,T)=f_1(R)+f_2(T)$, one finds $\nabla_\mu f_T=f_2''(T)\,\nabla_\mu T$, with no direct $\nabla R$ contribution. In either case, however, the presence of $f_R\neq \mathrm{const}$ in the field equations already introduces an additional scalar gravitational mode and higher-order derivatives of the metric. This propagating scalar degree of freedom persists even in vacuum, and is at the root of the severe phenomenological constraints on $f(R,T)$ models. As a result, detailed studies of cosmological perturbations and energy conditions \cite{Moraes2016,Shabani2013,Odintsov2013} have shown that the viable parameter windows in these theories are often very narrow and strongly model--dependent.

In $f(R,\mathcal L_m)$ gravity, proposed in \cite{Bertolami2007}, the modification takes the form of a single function of both $R$ and the matter Lagrangian, 
\begin{equation}
S=\int f(R,\mathcal L_m)\,\sqrt{-g}\,d^4x.
\end{equation}
The conservation law becomes
\begin{equation}
\nabla^\mu T_{\mu\nu}
=\frac{1}{f_{\mathcal L_m}}\,\big(\nabla^\mu f_{\mathcal L_m}\big)\big(g_{\mu\nu}\mathcal L_m-T_{\mu\nu}\big),
\end{equation}
with $f_{\mathcal L_m}=\partial f/\partial \mathcal L_m$. 
In general one has
\begin{equation}
\nabla_\mu f_{\mathcal L_m} 
= f_{\mathcal L_m R}\,\nabla_\mu R 
+ f_{\mathcal L_m \mathcal L_m}\,\nabla_\mu \mathcal L_m,
\end{equation}
so in the non--separable case $f_{\mathcal L_m}=f_{\mathcal L_m}(R,\mathcal L_m)$ curvature gradients explicitly enter the nonconservation law. 
For separable models of the form $f(R,\mathcal L_m)=f_1(R)+f_2(\mathcal L_m)$, one instead finds $\nabla_\mu f_{\mathcal L_m}=f_2''(\mathcal L_m)\,\nabla_\mu \mathcal L_m$, without a direct $\nabla R$ contribution. 
In either case, however, matter does not follow geodesics of the background metric, since the right-hand side of the conservation law induces an extra force proportional to $\nabla_\mu f_{\mathcal L_m}$. 
As in the $f(R,T)$ case, this feature severely restricts the allowed functional forms of the theory, as extensively discussed in \cite{Bertolami2008,Harko2010}.
The situation is different in the present $\phi(R,T)\mathcal L_m$ construction. The Einstein--Hilbert term is left untouched, and the modification enters solely through the matter weighting factor $\phi(R,T)$. In the case of interest $\phi=\phi(T)$, one has $\phi_R=0$, and the nonconservation law takes the form
\begin{equation}
\phi\,\nabla^\mu T_{\mu\nu}
= -(\nabla^\mu\phi)\,T_{\mu\nu}
+2\,\nabla^\mu\!\Big[\phi_T\,\mathcal L_m\,(T_{\mu\nu}+\Theta_{\mu\nu})\Big],
\end{equation}
so the departure from $\nabla^\mu T_{\mu\nu}=0$ is controlled entirely by gradients of matter quantities ($\nabla T$, $\nabla \mathcal L_m$) and not by curvature. Even if one allows $\phi=\phi(R,T)$, all the terms proportional to $\phi_R$ are multiplied by the matter Lagrangian and therefore vanish in vacuum. As a consequence, the Einstein equations are exactly recovered in empty space, with no extra scalar degree of freedom propagating in the geometry.  

This is the key structural difference: in $\phi(R,T)\mathcal L_m$ the nonconservation is a matter--weighting effect that disappears in vacuum, while in $f(R,T)$ and $f(R,\mathcal L_m)$ it is anchored in the geometric sector and persists even in the absence of matter. This distinction is emphasized visually in Fig.~\ref{fig:flow}, which shows how the three classes of theories descend from the Einstein--Hilbert baseline, and in Fig.~\ref{fig:origin-noncons}, which illustrates the different origins of the nonconservation law. Taken together, these comparisons highlight the milder and more controlled character of nonconservation in the $\phi(R,T)\mathcal L_m$ theory. The absence of extra geometric degrees of freedom in vacuum and the confinement of the violation to matter gradients explain why the stability window found in our cosmological analysis is broad and uniform, in contrast with the narrow and highly model--dependent domains of the other theories. A more technical discussion of the role of curvature gradients in the nonconservation law, 
and a detailed comparison between $\phi(R,T)\mathcal L_m$ and $f(R,\mathcal L_m)$ gravity, 
is provided in Appendix~\ref{app:noncons}.

Finally, it is important to clarify the relation with the general $f(R,\mathcal L_m,T)$ class proposed in \cite{HaghaniHarko2021}. 
Our construction is in fact contained within that framework: choosing 
$f(R,\mathcal L_m,T)=R+2\kappa(\phi(R,T)-1)\mathcal L_m$ in their action reproduces exactly 
the $\tfrac{1}{2\kappa}R+\phi(R,T)\mathcal L_m$ form considered here. 
In this sense, the present model should be regarded as a well–defined subcase of $f(R,\mathcal L_m,T)$ gravity. 
However, this particular branch has not been studied in detail before. 
By restricting to $\phi=\phi(T)$ with the infrared limit $\phi\to 1$, we show that it admits 
a stable and nonsingular cosmological bounce, a wide stability window, and a natural effective–field–theory interpretation 
as a density–dependent renormalization of the matter sector. 
These features distinguish our analysis from the broader but more formal setting of \cite{HaghaniHarko2021}, 
and justify treating $\phi(R,T)\mathcal L_m$ gravity as an interesting sector in its own right.

\begin{figure}[htbp]
\centering
\begin{tikzpicture}[
  box/.style={
    rectangle, draw, rounded corners, align=center, font=\small,
    minimum width=4.2cm, minimum height=1.2cm,
    top color=white, bottom color=#1!20,
    drop shadow, 
  },
  arrow/.style={-Latex, thick}
]

\node[box=gray] (gr) {\textbf{Einstein--Hilbert (GR)}\\ $R/2\kappa + \mathcal L_m$\\ Baseline};

\matrix (m) [below=2.5cm of gr, column sep=1cm, row sep=1cm] {
  \node[box=orange] (frt) {\textbf{$f(R,T)$ gravity}\\ Geometry modified\\ Explicit $T$ dependence}; &
  \node[box=green] (frlm) {\textbf{$f(R,\mathcal L_m)$ gravity}\\ Mixed $R+\mathcal L_m$\\ Curvature--matter coupling}; &
  \node[box=yellow] (phi) {\textbf{$\phi(R,T)\mathcal L_m$ (this work)}\\ Matter weighted by $\phi(T)$\\ Geometry = Einstein}; \\
};

\draw[arrow] (gr.south) -| (frt.north);
\draw[arrow] (gr.south) -- (frlm.north);
\draw[arrow] (gr.south) -| (phi.north);

\end{tikzpicture}
\caption{Schematic relation between GR and different modified gravity frameworks. 
From the Einstein--Hilbert baseline (top), one may either modify the geometric sector as in $f(R,T)$ gravity, 
introduce mixed curvature--matter couplings as in $f(R,\mathcal L_m)$ gravity, 
or keep Einstein geometry untouched while weighting the matter sector by $\phi(T)$, as in the present proposal.}
\label{fig:flow}
\end{figure}


\begin{figure}[htbp]
\centering
\begin{tikzpicture}[
  box/.style={
    rectangle, draw, rounded corners, align=center, font=\small,
    minimum width=4.2cm, minimum height=1.2cm,
    top color=white, bottom color=#1!20,
    drop shadow
  },
  lbl/.style={font=\footnotesize, align=center, text width=4.2cm},
  arrow/.style={-Latex, thick}
]

\node[box=gray] (gr) {\textbf{Einstein--Hilbert (GR)}\\
$R/2\kappa + \mathcal L_m$\\
$\nabla^\mu T_{\mu\nu}=0$};

\matrix (m) [below=2.6cm of gr, xshift=-1.0cm,
             column sep=0.8cm, row sep=1cm] {
  \node[box=orange] (frt) {\textbf{$f(R,T)$ gravity}\\ Action: $f(R,T)+\mathcal L_m$}; &
  \node[box=green] (frlm) {\textbf{$f(R,\mathcal L_m)$ gravity}\\ Action: $f(R,\mathcal L_m)$}; &
  \node[box=yellow] (phi) {\textbf{$\phi(R,T)\mathcal L_m$ (this work)}\\ Action: $R/2\kappa + \phi(R,T)\mathcal L_m$}; \\
};

\draw[arrow] (gr.south west) -- (frt.north);
\draw[arrow] (gr.south)      -- (frlm.north);
\draw[arrow] (gr.south east) -- (phi.north);

\node[lbl, below=0.35cm of frt]
{\(\nabla^\mu T_{\mu\nu}\neq 0\)\\ from \textbf{geometric} modification};
\node[lbl, below=0.35cm of frlm]
{\(\nabla^\mu T_{\mu\nu}\neq 0\)\\ from \textbf{curvature--matter} mixing};
\node[lbl, below=0.35cm of phi]
{\(\nabla^\mu T_{\mu\nu}\neq 0\)\\ from \textbf{matter weighting} $\phi(T)$};

\end{tikzpicture}
\caption{Conceptual origin of nonconservation in different frameworks. 
In GR the standard conservation law holds. 
In $f(R,T)$ and $f(R,\mathcal L_m)$ the violation of $\nabla^\mu T_{\mu\nu}=0$ is tied to the geometric sector, 
while in the present model it arises solely from the matter weighting $\phi(T)$, with the Einstein--Hilbert geometry left intact.}
\label{fig:origin-noncons}
\end{figure}
\paragraph{Relation to auxiliary-field constructions.}
The present framework is also related in spirit to theories in which
modifications of gravity arise from auxiliary (nondynamical) fields,
which can be eliminated algebraically at the level of the field equations
\cite{PaniSotiriouVernieri2013}. In such approaches, the resulting theory
can be written as general relativity supplemented by additional
divergence-free terms constructed from the stress--energy tensor and its
derivatives.

A key difference, however, is that in these constructions the matter sector
remains minimally coupled and satisfies $\nabla^\mu T_{\mu\nu}=0$, while
the modifications appear effectively on the geometric side of the field
equations. In contrast, in the present model the Einstein--Hilbert
geometric sector is kept exactly unchanged, and the modification is
implemented directly through a density-dependent weighting of the matter
Lagrangian. As a result, the nonconservation of $T_{\mu\nu}$ arises from
the matter sector itself rather than from an effective geometric
redefinition.

\subsection*{Semiclassical motivation and EFT interpretation}

A natural way to interpret the prefactor $\phi(T)$ is through the lens of semiclassical gravity and effective field theory (EFT). In semiclassical gravity the field equations take the form
\begin{equation}
G_{\mu\nu}+\Lambda g_{\mu\nu}=\kappa\,\langle \hat T_{\mu\nu}\rangle ,
\label{eq:semiclassical}
\end{equation}
and the one--loop effective action obtained after integrating out quantum matter fields generates local curvature invariants and matter--curvature operators \cite{BirrellDavies,ParkerToms,Donoghue,BOS,BarvinskyVilkovisky}. Schematically,
\begin{equation}
\Gamma_{\rm eff}[g,\Psi]
=\int\!\sqrt{-g}\,\Big[
\frac{1}{2\kappa}R
+\alpha_1 R^2+\alpha_2 R_{\mu\nu}R^{\mu\nu}
+\beta_1 R\,\mathcal O_m+\beta_2 \mathcal O_m^2
+\cdots\Big],
\label{eq:Geff}
\end{equation}
where $\Psi$ denotes matter, $\mathcal O_m$ are local scalar operators built from matter fields (and, in hydrodynamic limits, from fluid variables), and the Wilson coefficients $(\alpha_i,\beta_i)$ are suppressed by the heavy mass scales of the UV completion. A familiar example is the nonminimal operator $\xi R\phi^2$ for a scalar $\phi$; more generally, loop effects produce composite operators that couple curvature to matter bilinears and to the stress tensor \cite{Donoghue,ParkerToms}.

In hydrodynamic/effective descriptions of matter, the only covariant scalar at leading order is the trace $T\equiv g^{\mu\nu}T_{\mu\nu}$ (for a barotropic perfect fluid $T=-\rho+3p$). Thus, the leading EFT deformations involving the matter sector can be organized as a derivative expansion in powers of $T$ and its derivatives, e.g.
\begin{equation}
\Delta\mathcal L_{\rm eff}\;=\;\frac{c_1}{M^4}\,T\,\mathcal L_m
+\frac{c_2}{M^8}\,T^2\,\mathcal L_m
+\frac{d_1}{M^6}\,\nabla_\alpha T\,\nabla^\alpha T
+\cdots,
\label{eq:LeffT}
\end{equation}
with $M$ the EFT cutoff. Expanding a smooth $\phi(T)$ as a low--energy series,
\begin{equation}
\phi(T)\,\mathcal L_m=\mathcal L_m
+\frac{\beta_1}{M^4}\,T\,\mathcal L_m
+\frac{\beta_2}{M^8}\,T^2\,\mathcal L_m+\cdots,
\end{equation}
one precisely reproduces the \emph{non-derivative} subset of~\eqref{eq:LeffT}. Derivative operators such as $(\nabla T)^2$ are independent EFT terms, expected on general grounds and generically generated by radiative corrections. Their Wilson coefficients are unrelated to the Taylor coefficients $\beta_i$ unless a specific UV completion imposes matching relations. In this sense, the $\phi(T)$ construction implements an EFT--inspired parametrization that resums a tower of local $T^n$ operators, while remaining compatible with the existence of additional derivative terms in the full effective action.

This picture dovetails with the known structure of the one--loop effective action in curved spacetime. Matter loops generate local curvature terms ($R^2$, $R_{\mu\nu}R^{\mu\nu}$) and nonminimal couplings (e.g. $\xi R\phi^2$) \cite{BirrellDavies,BOS}, as well as nonlocal form factors capturing long--distance propagation \cite{BarvinskyVilkovisky,Donoghue}. Our construction deliberately \emph{does not} add higher--curvature geometric operators to the Einstein--Hilbert sector, thereby avoiding the extra dynamical scalar mode of $f(R)$--type models; instead, it parameterizes the dominant high--density corrections through $T$--dependent weights in the matter Lagrangian. This yields a controlled IR limit ($\phi\!\to\!1$) and a UV--agnostic\footnote{By ``UV--agnostic'' we mean that the framework does not commit to a particular ultraviolet completion of gravity (such as string theory or loop quantum gravity). Instead, it parametrizes possible high--density effects in a way consistent with effective field theory expectations, namely as an expansion in local stress--tensor operators.} but EFT--compatible high--density behavior, in which the bounce arises when the effective combination sourcing $H^2$ changes sign due to the $T$--dependent dressing.

Finally, we note that trace--anomaly effects provide an orthogonal, geometric source of quantum corrections of the form
\begin{equation}
\langle T^\mu{}_\mu\rangle_{\rm anom}=\frac{1}{(4\pi)^2}\big(c\,C_{\alpha\beta\gamma\delta}^2-a\,E+b\,\Box R\big),
\end{equation}
where $C^2$ is the Weyl tensor squared and $E$ the Euler density \cite{BirrellDavies,ParkerToms}. While our $\phi(T)\mathcal L_m$ model does not incorporate these purely geometric terms, it is \emph{complementary} to them: the anomaly corrects the geometric sector, whereas $\phi(T)$ captures matter--sector renormalization. Both are expected in a generic semiclassical effective action; here we isolate the latter to obtain a nonsingular, stable bounce without introducing higher--derivative geometric dynamics.
\section{A nonsingular, stable cosmological model}
In this section, a stable, nonsingular solution of the theory is constructed. As the matter sources, we consider a perfect fluid with a stress-energy tensor given by
\begin{equation}
T_{\mu\nu}=(p+\rho)u_{\mu}u_{\nu}+pg_{\mu\nu}
\end{equation}
Where $u^{\mu}$ is the four-velocity vector field satisfying the condition $u^{\mu}u_{\mu}=-1$. The matter Lagrangian acquires the standard form $\mathcal{L}_{m}=p$, and the tensor $\Theta_{\mu\nu}$ will be
\begin{equation}
\Theta_{\mu\nu}=-2T_{\mu\nu}+pg_{\mu\nu},
\end{equation}

\noindent and given the FLRW metric

\begin{equation}
    ds^2 = -dt^2 + a(t)^2\left(\frac{dr^2}{1 - Kr^2}+r^2d\Omega^2\right),
\end{equation}

We now apply the field equations to the case $\phi(R,T)\equiv\phi(T)$ choosing the nonlinear ansatz
\begin{equation}
    \phi(T) = 1 + \frac{\beta T}{1+\gamma T},
\end{equation}
where $\beta,\gamma$ are constants. For a barotropic fluid $p=w\rho$ we have $T=\tau\rho$, with
\begin{equation}
    \tau \equiv -1+3w.
\end{equation}
The effective fluid quantities sourcing Einstein’s equations take the form
\begin{equation}
    \rho_{\mathrm{eff}} = M(\rho)\,\rho, 
    \qquad p_{\mathrm{eff}}=\phi(\rho)\,w\,\rho,
\end{equation}
with
\begin{equation}
    M(\rho)=\phi(\rho)+2\phi_T(\rho)\,w(1+w)\,\rho,
    \qquad
    A(\rho)=\phi(\rho)+2\phi_T(\rho)\,w\,\rho.
\end{equation}
Here
\begin{equation}
    \phi(\rho) = 1+\frac{\beta \tau\rho}{1+\gamma\tau\rho},
    \qquad
    \phi_T(\rho) = \frac{\beta}{\big(1+\gamma\tau\rho\big)^2},
    \qquad
    \frac{d\phi_T}{d\rho}=-\frac{2\beta\gamma\tau}{\big(1+\gamma\tau\rho\big)^3}.
\end{equation}
\paragraph{On the parameters of the rational ansatz.}
The parameters $\beta$ and $\gamma$ appearing in the function
\[
\phi(T)=1+\frac{\beta\,T}{1+\gamma\,T} ,
\]
have dimensions $[\beta]=[\gamma]=M^{-4}$, since the trace $T$ has the same
dimensions as the energy density.

They therefore define a characteristic energy scale $M$ at which the
modifications become relevant, with $M^4\sim 1/\beta$ and $M^4\sim 1/\gamma$.
In the low-density regime, the ansatz admits the expansion
\begin{equation}
\phi(T)=1+\beta T + \mathcal{O}(T^2),
\end{equation}
which can be interpreted as the leading-order correction in an effective
field theory expansion.

The rational form can be viewed as a resummation that ensures a controlled
behavior at high densities, avoiding divergences and allowing for a smooth
interpolation between the infrared limit $\phi\to1$ and the nonlinear regime.
In this sense, $\beta$ controls the strength of the correction, while $\gamma$
regulates its saturation at high densities.

\subsection{Modified Friedmann equations}

Since the geometric part of the action is Einstein–Hilbert, the background equations keep their GR form with $\rho_{\mathrm{eff}},p_{\mathrm{eff}}$:

\begin{equation}\label{Friedmann}
H^2 = \frac{\kappa}{3}\,M(\rho)\,\rho,
\end{equation}
\begin{equation}\label{Raychaudhuri}
\dot H = -\frac{\kappa}{2}\,(1+w)\,\rho\,A(\rho),
\end{equation}
\begin{equation}\label{Continuity}
\dot\rho = -3H(1+w)\rho \;\frac{A(\rho)}{N(\rho)},
\qquad N(\rho)\equiv M(\rho)+\rho M'(\rho).
\end{equation}

The derivative $M'(\rho)=dM/d\rho$ evaluates to
\begin{equation}
M'(\rho)=\frac{\beta\tau}{(1+\gamma\tau\rho)^2}
+2w(1+w)\left[\frac{\beta}{(1+\gamma\tau\rho)^2}
-\frac{2\beta\gamma\tau\rho}{(1+\gamma\tau\rho)^3}\right].
\end{equation}

\subsection{Bounce density and smoothness}

The bounce occurs when $M(\rho_b)=0$, i.e. $H(\rho_b)=0$.  
With $p=w\rho$ and $\tau\equiv -1+3w$, defining $x\equiv\tau\rho$, the bounce condition becomes
\begin{equation}
\big(\gamma^2+\beta\gamma\big)x^2
+\left(2\gamma+\beta+\frac{2\beta\,w(1+w)}{\tau}\right)x
+1=0, \qquad x=\tau\rho.
\label{eq:bounce_quadratic}
\end{equation}
This is a quadratic equation in $x$, subject to the regularity condition $1+\gamma x\neq 0$ to exclude the pole of $\phi_T$.

\paragraph{Explicit solution.}  
Let
\begin{equation}
a=\gamma(\gamma+\beta),\quad
b=2\gamma+\beta+\frac{2\beta w(1+w)}{\tau},\quad
c=1,
\end{equation}
with discriminant $\Delta=b^2-4ac$. Then the solutions are
\begin{equation}
x_b=\frac{-\,b\pm\sqrt{\Delta}}{2a},\qquad
\rho_b=\frac{x_b}{\tau}>0,\qquad 1+\gamma x_b\neq 0.
\end{equation}
A physical bounce requires $\Delta\geq 0$ and a positive root $\rho_b$.

\paragraph{Smoothness condition.}  
The Raychaudhuri equation,
\begin{equation}
\dot H=-\frac{\kappa}{2}(1+w)\,\rho\,A(\rho),
\end{equation}
evaluates at the bounce to
\begin{equation}
A(\rho_b)=-\,\frac{2\,\beta\,w^2\,\rho_b}{\big(1+\gamma\tau\rho_b\big)^2}.
\end{equation}
Therefore,
\begin{equation}
\dot H(\rho_b)
=-\frac{\kappa}{2}(1+w)\rho_b A(\rho_b)
=\frac{\kappa\,\beta\,w^2(1+w)\,\rho_b^2}{\big(1+\gamma\tau\rho_b\big)^2}.
\label{eq:hdot_bounce}
\end{equation}

\paragraph{Result.}  
For $w\neq 0$ and $\beta>0$ one has $\dot H(\rho_b)>0$, ensuring that the bounce is smooth and nonsingular.  
The case $w=0$ (dust) gives $A(\rho_b)=0$ and hence $\dot H(\rho_b)=0$, a marginal case consistent with $\mathcal L_m=p=0$ where the modification is absent. The physical root $x_b$ with $\rho_b=x_b/\tau>0$ determines the bounce density.  More precisely, smoothness requires
\begin{equation}
\dot H(\rho_b) = -\frac{\kappa}{2}(1+w)\rho_b A(\rho_b) > 0,
\end{equation}
i.e. the function $A$ must verify $A(\rho_b)<0$ for $w>-1$.

\subsection{Energy conditions near the bounce.}

The occurrence of a cosmological bounce requires a violation of the
null energy condition (NEC) at the level of the effective dynamics,
i.e.
\begin{equation}
\rho_{\rm eff} + p_{\rm eff} < 0.
\end{equation}
In the present framework, this violation arises from the
$\phi(T)$-dependent modification of the matter sector, which alters
the effective energy--momentum tensor while leaving the underlying
matter fields unchanged.

For completeness, the effective energy conditions can be written as
\begin{align}
\mathrm{WEC:}\quad & \rho_{\rm eff} = M(\rho)\,\rho \geq 0, \\
\mathrm{NEC:}\quad & \rho_{\rm eff} + p_{\rm eff}
= (1+w)\,\rho\,A(\rho) \geq 0, \\
\mathrm{SEC:}\quad & \rho_{\rm eff} + 3p_{\rm eff}
= M(\rho)\,\rho + 3\phi\,w\rho \geq 0, \\
\mathrm{DEC:}\quad & \rho_{\rm eff} \geq |p_{\rm eff}|
= \phi\,|w|\,\rho.
\end{align}

\begin{figure}[h]
    \centering
    \includegraphics[width=\textwidth]{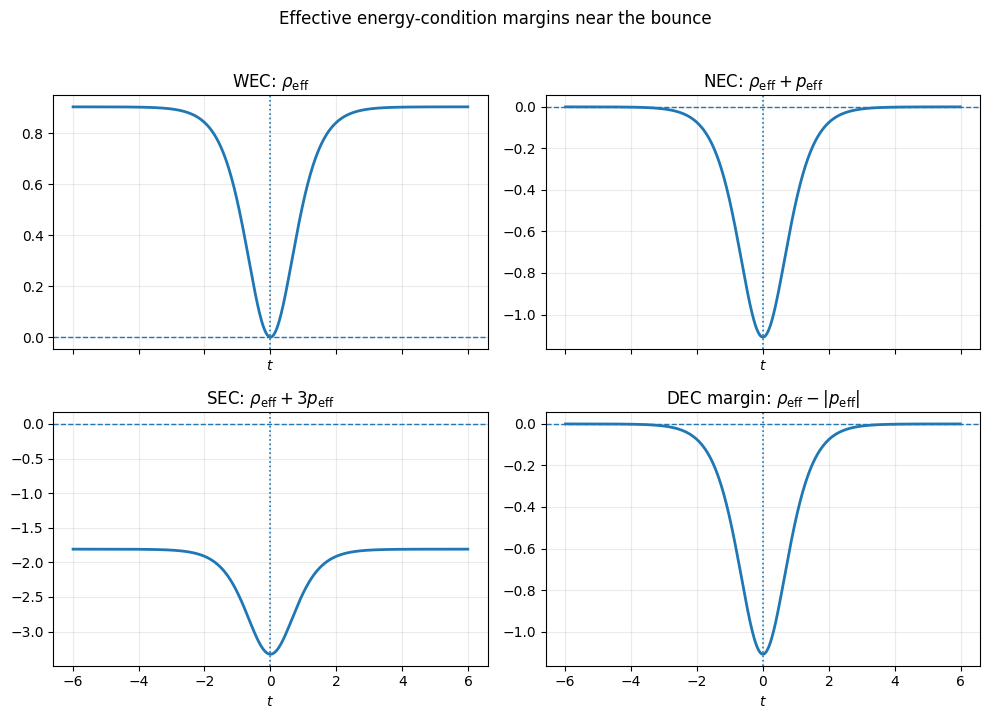}
    \caption{
Effective energy-condition margins near the bounce.
The WEC is saturated at the bounce, while the NEC and SEC are
violated in its vicinity. The DEC is also violated in a small
neighborhood around the bounce. The vertical dashed line marks
the bounce.
}
    \label{fig:energy}
\end{figure}

The behavior of these quantities in the vicinity of the bounce is
displayed in Fig.~\ref{fig:energy}.
One observes that the weak energy condition (WEC) is saturated at the
bounce, where $\rho_{\rm eff}=0$, while the null and strong energy
conditions (NEC and SEC) become negative in a finite neighborhood,
providing the necessary departure from standard general relativity to
enable a nonsingular transition.

The dominant energy condition (DEC) is also seen to be violated in a
small region around the bounce. This follows from the fact that
$\rho_{\rm eff}=0$ at the bounce, while $p_{\rm eff}$ remains finite
for $w\neq 0$. However, this violation is confined to the effective
description and arises from the nonminimal coupling encoded in
$\phi(T)$, while the underlying matter sector may still satisfy the
standard energy conditions.

Away from the bounce, all effective energy conditions are rapidly
restored, and the system approaches a regime consistent with standard
cosmological evolution.

\begin{figure}[h]
    \centering
    \includegraphics[width=0.75\textwidth]{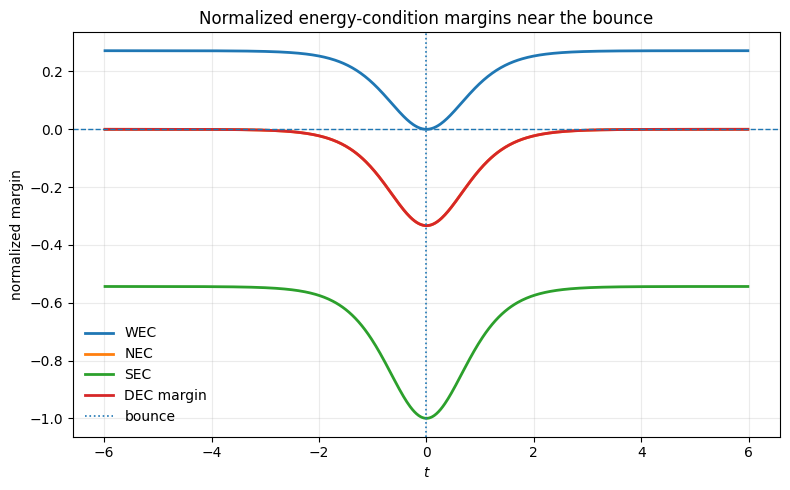}
   \caption{
Normalized energy-condition margins near the bounce.
The NEC is strongly suppressed relative to the WEC, indicating
$\rho_{\rm eff}+p_{\rm eff}\approx 0$ and an effective equation
of state $w_{\rm eff}\simeq -1$ in the bounce region.
}
    \label{fig:energy_norm}
\end{figure}

To better visualize the relative hierarchy between the different
conditions, we show in Fig.~\ref{fig:energy_norm} the corresponding
normalized quantities. In this representation, the NEC appears strongly
suppressed with respect to the WEC in the vicinity of the bounce,
indicating that
\begin{equation}
\rho_{\rm eff}+p_{\rm eff} \approx 0.
\end{equation}
This corresponds to an effective equation of state approaching a
vacuum-like regime,
\begin{equation}
w_{\rm eff} \equiv \frac{p_{\rm eff}}{\rho_{\rm eff}} \simeq -1,
\end{equation}
in the neighborhood of the bounce. Strictly at the bounce,
$\rho_{\rm eff}=0$ and $w_{\rm eff}$ is not defined, but the system
smoothly approaches this limit.

This behavior provides a clear physical interpretation of the bounce:
the modified coupling induces an effective fluid that transiently
mimics a cosmological-constant-like phase, allowing for a smooth
nonsingular transition without introducing exotic fundamental matter
components.

\subsection{Cosmological dynamics}

The background evolution is obtained by numerically integrating
Eqs.~(\ref{Friedmann})--(\ref{Continuity}). The system is evolved
in terms of the variables $(\rho,a)$, while the Hubble parameter
is reconstructed from the Friedmann equation,
\begin{equation}
H = \pm \sqrt{\frac{\kappa}{3} M(\rho)\,\rho},
\end{equation}
where the sign distinguishes the contracting ($H<0$) and expanding
($H>0$) branches.

The integration is initialized slightly away from the bounce density,
$\rho=\rho_b(1-\epsilon)$ with $\epsilon\ll 1$, in order to avoid
the degeneracy of the equations at $H=0$. The contracting and
expanding solutions are then reconstructed symmetrically by matching
the two branches at the bounce.

Numerical integration shows that:
\begin{itemize}
    \item The scale factor $a(t)$ reaches a finite minimum at $t_b$,
    where $H=0$.
    \item The Hubble parameter crosses zero with finite positive slope
    $\dot H(t_b)>0$, confirming a smooth bounce.
    \item The energy density $\rho(t)$ remains bounded and reaches a
    finite maximum at the bounce.
\end{itemize}

\begin{figure}[htbp]
    \centering
    \includegraphics[width=\textwidth]{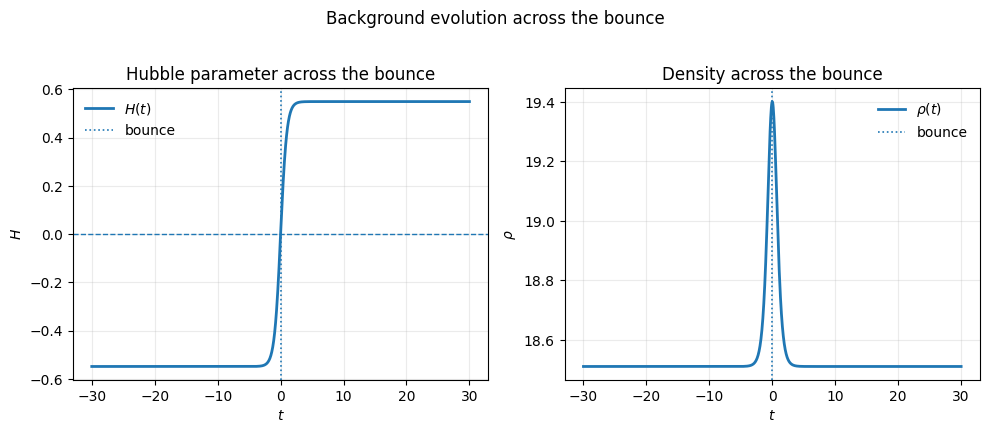}
    \caption{Background evolution across the bounce. The Hubble parameter smoothly
changes sign from contraction to expansion, while the matter density
reaches a finite maximum at the bounce. The vertical dotted line marks
the bounce.}
    \label{fig:allplots}
\end{figure}

These features are illustrated in Fig.~\ref{fig:allplots},
which displays the evolution of $H(t)$ and $\rho(t)$ for representative
parameters $(w=0.1, \beta=0.1, \gamma=0.01)$. The transition from
contraction to expansion occurs smoothly, and all physical quantities
remain finite throughout the evolution.

It is also instructive to quote explicit numerical values for the
bounce quantities. For the above parameters, the quadratic
condition~\eqref{eq:bounce_quadratic} yields the first positive root
\begin{equation}
\rho_b \simeq 19.4\,\rho_c,
\end{equation}
where $\rho_c=\kappa^{-1}$ defines the characteristic density scale.
Evaluating Eq.~\eqref{eq:hdot_bounce} at this value gives
\begin{equation}
\dot H(\rho_b)\simeq 0.55\,\kappa \rho_c,
\end{equation}
which is finite and strictly positive, confirming that the bounce is
smooth.
At late times, the energy density decreases monotonically and the
system approaches the standard low-density regime, where
$\phi(T)\to 1$ and general relativity is recovered.

These results demonstrate quantitatively that the
$\phi(T)\mathcal L_m$ framework regulates the high-density regime
without introducing divergences. The scale factor never vanishes,
the density remains bounded by $\rho_b$, and the Hubble parameter
evolves smoothly across the bounce. The transition from contraction
to expansion is therefore dynamically controlled by the modified
matter coupling.

\subsection{Stability analysis}
Beyond the background dynamics, it is important to test the internal stability of the effective fluid description. Two useful quantities are the function $Q(t)$, which quantifies deviations from the GR continuity law, and the effective sound speed squared $c_s^2$, which controls the stability of scalar perturbations. 

\begin{figure}[htbp]
    \centering
    \includegraphics[width=\textwidth]{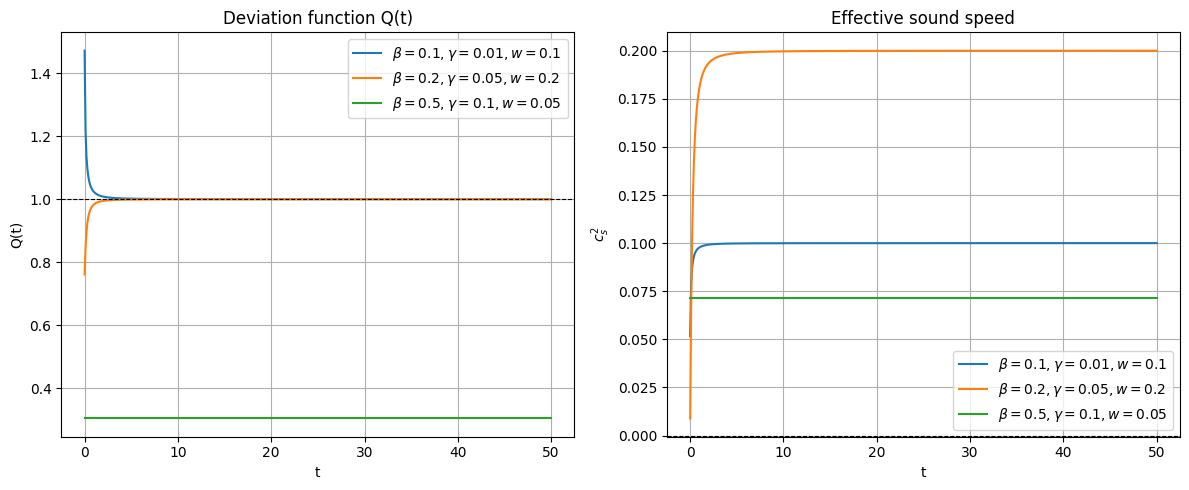}
    \caption{Stability diagnostics of the $\phi(T)\mathcal L_m$ model. 
    \textbf{Left:} the deviation function $Q(t)=A(\rho)/N(\rho)$, which quantifies the departure from the GR continuity law. 
    Near the bounce $Q(t)\neq 1$, signaling matter--geometry energy exchange, while at late times $Q\to 1$, recovering GR. 
    \textbf{Right:} the effective sound speed squared $c_s^2(t)=\dot p_{\rm eff}/\dot \rho_{\rm eff}$. 
    In the physical parameter window $c_s^2$ remains positive (and close to $w$), ensuring perturbative stability across the bounce.}
    \label{fig:Qcs2}
\end{figure}

\paragraph{Continuity deformation $Q(t)$.}
From the modified continuity equation
\begin{equation}
\dot\rho = -\,3H(1+w)\rho\,\frac{A(\rho)}{N(\rho)},
\qquad
N(\rho)\equiv M(\rho)+\rho M'(\rho),
\end{equation}
we define
\begin{equation}
Q(t)\equiv \frac{A(\rho)}{N(\rho)}.
\end{equation}
In GR one has $Q=1$. 
In the present model, $Q(t)$ departs from unity in the high-density regime near the bounce, signaling an effective energy exchange between the matter and the geometric sector. 
At late times $Q\to 1$, and the standard continuity equation of GR is recovered. 
This behavior is illustrated in Fig.~\ref{fig:Qcs2}.

\paragraph{Effective sound speed $c_s^2$.}
The effective sound speed is defined as
\begin{equation}
c_s^2 \equiv \frac{\dot p_{\rm eff}}{\dot \rho_{\rm eff}},
\end{equation}
with
\begin{equation}
\rho_{\rm eff}=M(\rho)\rho,\qquad p_{\rm eff}=\phi(\rho)\,w\rho.
\end{equation}
A direct calculation yields
\begin{equation}
c_s^2=\frac{w\Big[\phi(\rho)+\rho\,\phi'(\rho)\Big]}{M(\rho)+\rho\,M'(\rho)}.
\end{equation}
Positivity of $c_s^2$ guarantees stability of scalar perturbations, while $c_s^2\leq 1$ avoids superluminal propagation. 
Numerical evaluation shows that $c_s^2(t)$ remains positive and subluminal across the bounce for the allowed window of parameters, confirming the dynamical stability of the model (Fig.~\ref{fig:Qcs2}).

\subsubsection*{IR consistency: $Q(\rho)\to 1$ as $\rho\to 0$}

Recall
\[
Q(\rho)\equiv \frac{A(\rho)}{N(\rho)},\qquad
A(\rho)=\phi(\rho)+2\phi_T(\rho)\,w\,\rho,\qquad
N(\rho)=M(\rho)+\rho M'(\rho),
\]
with
\[
M(\rho)=\phi(\rho)+2\phi_T(\rho)\,w(1+w)\,\rho,\qquad
\phi(\rho)=1+\frac{\beta\,\tau\rho}{1+\gamma\tau\rho},\qquad
\phi_T(\rho)=\frac{\beta}{\big(1+\gamma\tau\rho\big)^2},
\]
and \(\tau\equiv -1+3w\).

\paragraph{Exact IR value.}
At $\rho=0$ one has $\phi(0)=1$, $\phi_T(0)=\beta$, hence
\[
A(0)=\phi(0)+2\phi_T(0)w\cdot 0 = 1,\qquad
M(0)=\phi(0)+2\phi_T(0)w(1+w)\cdot 0 = 1.
\]
Therefore
\[
N(0)=M(0)+0\cdot M'(0)=1,
\qquad\Rightarrow\qquad
Q(0)=\frac{A(0)}{N(0)}=1.
\]

\paragraph{Small-$\rho$ expansion.}
For $|\gamma\tau\rho|\ll 1$,
\[
\phi(\rho)=1+\beta\tau\rho+\mathcal{O}(\rho^2),
\qquad
\phi_T(\rho)=\beta+\mathcal{O}(\rho),
\]
so
\[
A(\rho)=1+\beta(\tau+2w)\rho+\mathcal{O}(\rho^2),\qquad
M(\rho)=1+\beta\big(\tau+2w(1+w)\big)\rho+\mathcal{O}(\rho^2).
\]
Moreover,
\[
M'(0)=\beta\tau+2\beta w(1+w),
\quad\Rightarrow\quad
N(\rho)=M(\rho)+\rho M'(\rho)=1+2\beta\big(\tau+2w(1+w)\big)\rho+\mathcal{O}(\rho^2).
\]
Hence
\[
Q(\rho)=\frac{1+\beta(\tau+2w)\rho+\mathcal{O}(\rho^2)}{1+2\beta(\tau+2w(1+w))\rho+\mathcal{O}(\rho^2)}
=1+\big[\beta(\tau+2w)-2\beta(\tau+2w(1+w))\big]\rho+\mathcal{O}(\rho^2),
\]
which proves
\[
\boxed{\;\lim_{\rho\to 0}Q(\rho)=1\;}.
\]
\paragraph{Interpretation.}
Therefore the modified continuity law
$\dot\rho=-3H(1+w)\rho\,\frac{A}{N}$
reduces to the GR form $\dot\rho=-3H(1+w)\rho$ in the infrared, as expected.
Deviations $Q\neq 1$ are confined to the high-density regime near the bounce.
\subsubsection*{2D parameter stability analysis}

\begin{figure}[htbp]
    \centering
    \includegraphics[width=0.75\textwidth]{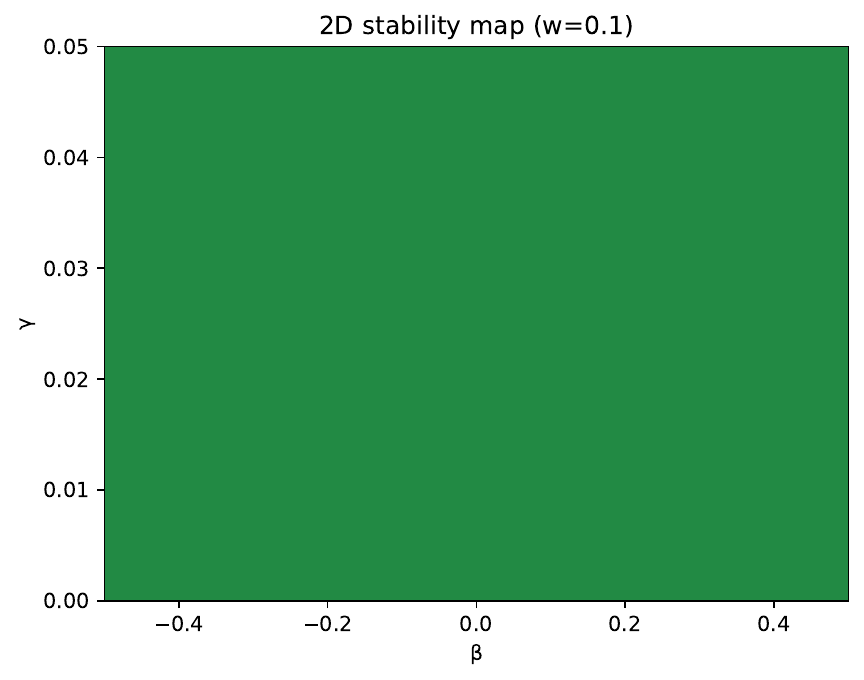}
    \caption{Two-dimensional stability map in the $(\beta,\gamma)$ plane for $w=0.1$. 
    The shaded region corresponds to stable models with $c_s^2>0$ and a regular $Q(\rho)$, ensuring that the bounce is both nonsingular and perturbatively stable. 
    Outside this region the theory develops gradient instabilities or singularities in $Q(\rho)$.}
    \label{fig:stab2d}
\end{figure}

In order to complement the 1D time evolution, it is useful to explore the stability conditions directly in the parameter space $(\beta,\gamma)$ for fixed $w$. 
For each point of the grid we evaluate the sign of the effective sound speed squared $c_s^2$ and the regularity of the function $Q(\rho)$. 
The resulting diagram identifies the regions where the model is stable and free of singularities. 
Figure~\ref{fig:stab2d} shows an example for $w=0.1$, with shaded regions indicating $c_s^2>0$ and $Q$ well-defined across the bounce.

The fact that the stability map in Fig.~\ref{fig:stab2d} is uniformly stable within the considered parameter window 
($-0.5 \leq \beta \leq 0.5$, $0 \leq \gamma \leq 0.05$) is a highly desirable feature of the model. 
It means that the existence of a smooth bounce and the absence of gradient instabilities do not rely on fine-tuned values of the coupling parameters. 
Instead, the theory possesses a robust domain of stability in which the energy conditions, the positivity of the effective sound speed squared, and the infrared consistency $Q\to 1$ are all simultaneously satisfied. 
This reinforces the claim that the proposed $\phi(T)\mathcal L_m$ modification provides a consistent and viable alternative to GR at high densities, while smoothly recovering standard cosmology at late times. Indeed, the combined stability analysis confirms that the $\phi(T)\mathcal L_m$ model is dynamically well behaved. 
In the one-dimensional time evolution (Fig.~\ref{fig:Qcs2}), the deviation function $Q(t)$ departs from unity only in the high-density regime and approaches $Q\to 1$ as $\rho\to 0$, ensuring infrared consistency, while the effective sound speed $c_s^2(t)$ remains positive and subluminal across the bounce. 
In the two-dimensional parameter scan (Fig.~\ref{fig:stab2d}), we find that the entire window $-0.5 \leq \beta \leq 0.5$, $0 \leq \gamma \leq 0.05$ is uniformly stable, with no signs of gradient instabilities or singular behavior. 
Taken together, these results show that the bounce solutions are not a fine-tuned artifact, but rather a generic prediction of the model within a broad and robust region of parameter space.

However, this analysis does not replace a full perturbative treatment.
A complete stability study would require the investigation of linear
cosmological perturbations, including scalar, vector, and tensor modes
within a gauge-invariant framework. This lies beyond the scope of the
present work and will be addressed in future studies.

The present results should therefore be interpreted as evidence of effective
consistency rather than a proof of full perturbative stability.
\paragraph{On the role of radiation and the trace $T$.}
For an ideal conformal radiation fluid with $w=1/3$, one has $T=-\rho+3p=0$,
and therefore $\phi(T)\to 1$, so that the modification effectively switches off.
This implies that the bounce mechanism discussed here is not sourced by a
perfectly conformal radiation component alone.

However, in realistic high-density regimes the cosmic medium is not expected
to be exactly described by a single ideal radiation fluid. Mixtures of components,
massive species, departures from equilibrium, viscous effects, or quantum
trace-anomaly contributions can all generate a nonvanishing effective trace.
In such situations, the $\phi(T)\mathcal L_m$ coupling remains active and can
drive the effective source term to zero at finite density, leading to a
nonsingular bounce.

The present results should therefore be interpreted as applying to
high-density regimes with nonvanishing effective trace, rather than to a
purely conformal radiation phase.
\subsection{Low-density limit and comparison with $\Lambda$CDM}

For $\rho \ll \rho_b$ one finds
\begin{equation}
\phi(\rho)\to 1, \quad \rho_{\rm eff}\to \rho, \quad p_{\rm eff}\to p,
\end{equation}
so that the standard Friedmann equations are recovered. Thus, the model reproduces the successful predictions of $\Lambda$CDM at late times and in the weak–field regime, while replacing the initial singularity with a smooth bounce at high densities.
 \section{Summary and Conclusions}

In this work, we have developed and analyzed a novel class of modified gravity theories in which the Einstein--Hilbert geometric sector remains unaltered, while the matter Lagrangian is weighted by a nontrivial function $\phi(R,T)$ of the Ricci scalar $R$ and the trace $T$ of the energy--momentum tensor. This construction differs structurally from the better-known $f(R,T)$ and $f(R,\mathcal L_m)$ frameworks: whereas the former modifies the geometric part of the action by replacing $R$ with a general function of $R$ and $T$, and the latter introduces explicit curvature--matter couplings, the present model leaves the geometry strictly Einsteinian and encodes the new physics exclusively in a trace-dependent weighting of the matter sector. As a consequence, no additional geometric degrees of freedom are introduced in vacuum, ensuring consistency with local gravity tests, while the coupling to matter is effectively renormalized in a density-dependent way.  

Focusing on the nonlinear ansatz $\phi(T)=1+\beta T/(1+\gamma T)$, we derived the modified Friedmann and Raychaudhuri equations, together with the effective continuity law. The system reduces to the standard GR equations in the low-density regime, thereby recovering $\Lambda$CDM phenomenology at late times, but departs substantially from GR at high densities. We have shown analytically that the condition $M(\rho_b)=0$ admits a finite positive solution for the bounce density $\rho_b$, and that the derivative $\dot H(\rho_b)$ is finite and positive provided $w\neq 0$ and $\beta>0$. These results guarantee the existence of a smooth, nonsingular cosmological bounce. The maximum density at the bounce is bounded, the Hubble parameter vanishes and changes sign continuously, and the scale factor reaches a finite minimum, thus resolving the big-bang singularity within a stable dynamical setting.  

We have further examined the stability properties of the model. The effective energy conditions indicate that the Weak and Dominant Energy Conditions remain satisfied, while the Null and Strong Energy Conditions are transiently relaxed in the vicinity of the bounce, precisely as required to permit a nonsingular transition. The continuity deformation function $Q(t)$ departs from unity only in the high-density regime and asymptotically approaches $Q\to 1$ at late times, ensuring infrared consistency with GR. In addition, the effective sound speed squared $c_s^2$ remains positive and subluminal across the entire evolution, confirming the perturbative stability of the model.  

Altogether, these findings establish that the $\phi(R,T)\mathcal L_m$ framework provides a consistent and physically viable modification of GR. It replaces the initial singularity with a smooth bounce, preserves stability of both background and perturbations, and reduces to the standard cosmological model in the low-density limit. The theory can be interpreted as an effective renormalization of the matter sector, with corrections that become relevant at high densities but fade away at late times. This suggests a natural path to singularity resolution without compromising the empirical successes of $\Lambda$CDM.  

Nevertheless, it is important to remain cautious about the scope of these results. While the present framework succeeds in resolving the big bang singularity in a controlled and technically consistent manner, it is not yet clear whether it can reproduce or replace the rich phenomenology that has been attributed to other theories of modified gravity, such as effective models of dark matter or dark energy. The simplicity of $\phi(R,T)\mathcal L_m$ may ultimately be a strength---by avoiding additional degrees of freedom and pathological instabilities---or it may turn out to be a limitation if the model lacks sufficient dynamical flexibility to account for the full range of cosmological observations.  

For these reasons, the present proposal should be regarded as a proof of principle: it demonstrates that it is possible to obtain a robust nonsingular bounce without modifying the geometric sector of GR, but further work is required to assess its viability in more general settings. In particular, the analysis of cosmological perturbations, the exploration of astrophysical applications, and the study of whether the framework can naturally incorporate dark sector phenomenology will be crucial steps in determining whether $\phi(R,T)\mathcal L_m$ represents a viable alternative to GR or merely a limited effective model.\\  
It should be emphasized that the present model belongs to the broader $f(R,\mathcal L_m,T)$ class introduced in \cite{HaghaniHarko2021}. 
While that general framework provides the umbrella for different curvature–matter couplings, the specific branch 
$f(R,\mathcal L_m,T)=R+2\kappa(\phi(R,T)-1)\mathcal L_m$ with $\phi(T)\!\to\!1$ in the infrared has not been investigated before. 
By focusing on this sector, we uncover robust phenomenological consequences — including a nonsingular bounce, a wide stability window, 
and a natural effective field theory interpretation — that are not apparent in the general treatment.

\section*{Acknowledgements}
The authors thank the anonymous referee for constructive and insightful
comments that helped to improve the quality of the manuscript.

\appendix
\section{Nonconservation laws and curvature gradients in $\phi(R,T)\mathcal L_m$ versus $f(R,\mathcal L_m$)}
\label{app:noncons}

In the present framework
\begin{equation}
S=\int \Big[\tfrac{1}{2\kappa}R+\phi(R,T)\,\mathcal L_m\Big]\sqrt{-g}\,d^4x,
\end{equation}
variation with respect to $g^{\mu\nu}$ yields an effective tensor
\begin{equation}
T^{\rm eff}_{\mu\nu}
=\phi\,T_{\mu\nu}
-2\,\phi_R\,\mathcal L_m\,R_{\mu\nu}
-2\Big(g_{\mu\nu}\Box-\nabla_\mu\nabla_\nu\Big)(\phi_R\,\mathcal L_m)
-2\,\phi_T\,\mathcal L_m\,(T_{\mu\nu}+\Theta_{\mu\nu}),
\label{eq:Teff-philm-app}
\end{equation}
where $\phi_R\equiv \partial\phi/\partial R$, $\phi_T\equiv \partial\phi/\partial T$, and $\Theta_{\mu\nu}\equiv g^{\alpha\beta}\delta T_{\alpha\beta}/\delta g^{\mu\nu}$. 
The Bianchi identity implies $\nabla^\mu T^{\rm eff}_{\mu\nu}=0$, which can be rearranged as a modified conservation law for $T_{\mu\nu}$:
\begin{align}
\phi\,\nabla^\mu T_{\mu\nu}
&= -(\nabla^\mu \phi)\,T_{\mu\nu}
+2\,\nabla^\mu\!\Big[\phi_T\,\mathcal L_m\,(T_{\mu\nu}+\Theta_{\mu\nu})\Big]
+2\,\phi_R\,\nabla^\mu(\mathcal L_m R_{\mu\nu})\nonumber\\
&\quad
+2\,\nabla^\mu\!\Big\{\big(g_{\mu\nu}\Box-\nabla_\mu\nabla_\nu\big)(\phi_R \mathcal L_m)\Big\}.
\label{eq:noncons-philm-app}
\end{align}
After expanding derivatives (using Bianchi identities and commutators), gradients of the curvature scalar do appear when $\phi_R\neq 0$, but \emph{always multiplied by $\mathcal L_m$ (or $\nabla\mathcal L_m$)}. 
Therefore, in vacuum ($\mathcal L_m=0=T_{\mu\nu}$) the right-hand side of \eqref{eq:noncons-philm-app} vanishes and the geometric sector is exactly Einsteinian; no extra propagating scalar mode survives in vacuum.

This structural feature contrasts with $f(R,\mathcal L_m)$ gravity,
\begin{equation}
S=\int f(R,\mathcal L_m)\,\sqrt{-g}\,d^4x,
\end{equation}
for which the conservation law reads \cite{Bertolami2007,Bertolami2008,Harko2010}
\begin{equation}
\nabla^\mu T_{\mu\nu}
=\frac{1}{f_{\mathcal L_m}}\,(\nabla^\mu f_{\mathcal L_m})\,(g_{\mu\nu}\mathcal L_m-T_{\mu\nu}),
\qquad
\nabla_\mu f_{\mathcal L_m}
= f_{\mathcal L_m R}\,\nabla_\mu R + f_{\mathcal L_m\mathcal L_m}\,\nabla_\mu\mathcal L_m,
\label{eq:noncons-fRLM-app}
\end{equation}
So curvature gradients enter the nonconservation law in the non--separable case ($f_{\mathcal L_m R}\neq 0$), 
where $\nabla_\mu f_{\mathcal L_m}$ contains $\nabla R$. 
In addition, and independently of this point, the field equations of $f(R,\mathcal L_m)$ contain the combination 
\[
f_R R_{\mu\nu} - \tfrac{1}{2}f g_{\mu\nu} + (g_{\mu\nu}\Box - \nabla_\mu\nabla_\nu)f_R,
\]
with $f_R\equiv \partial f/\partial R$. 
Whenever $f_R$ depends on $R$, the derivative terms $(\Box f_R, \nabla_\mu\nabla_\nu f_R)$ introduce higher derivatives of the metric, 
which correspond to an additional scalar gravitational degree of freedom. 
Unlike in the $\phi(R,T)\mathcal L_m$ case, this extra mode persists even in vacuum, since one generally has $f(R,0)\neq R/2\kappa$. 
Only by fine--tuning the functional form to $f(R,0)=R/2\kappa$ does the theory reduce to Einstein gravity in the absence of matter.
In summary:
\begin{itemize}
\item In $\phi(R,T)\mathcal L_m$, curvature-gradient effects in the force law are matter-suppressed: they vanish in vacuum and do not change the content of the geometric sector.
\item In $f(R,\mathcal L_m)$, the extra scalar mode belongs to the geometry itself and persists in the absence of matter.
\end{itemize}

This is the precise sense in which the present framework can be regarded as ``more benign'' than $f(R,\mathcal L_m)$: not because $\nabla R$ never appears, but because such terms are modulated by $\mathcal L_m$ and switch off in vacuum, where the theory reduces to pure GR.
\section{Scalar field realization of the $\phi(R,T)\mathcal L_m$ theory}
\label{app:scalarfield}

Consider a canonical scalar field with
\[
\mathcal L_m=-\tfrac12\,g^{\mu\nu}\partial_\mu\varphi\,\partial_\nu\varphi - V(\varphi),
\]
whose energy--momentum tensor is
\[
T_{\mu\nu}=\partial_\mu\varphi\,\partial_\nu\varphi
- g_{\mu\nu}\Big(\tfrac12(\partial\varphi)^2+V\Big),\qquad
T\equiv g^{\mu\nu}T_{\mu\nu}=(\partial\varphi)^2-4V(\varphi).
\]
The effective matter action is
\[
S_m^{\rm eff}=\int \sqrt{-g}\;\phi(T)\,\mathcal L_m.
\]

\subsection*{Modified Klein--Gordon equation}
Varying with respect to $\varphi$ and using
\[
\frac{\partial \mathcal L_m}{\partial(\nabla_\mu\varphi)}=-\nabla^\mu\varphi,\quad
\frac{\partial \mathcal L_m}{\partial \varphi}=-V_{,\varphi},\quad
\frac{\partial T}{\partial(\nabla_\mu\varphi)}=2\,\nabla^\mu\varphi,\quad
\frac{\partial T}{\partial \varphi}=-4\,V_{,\varphi},
\]
we obtain
\begin{align*}
\frac{\partial (\phi\,\mathcal L_m)}{\partial(\nabla_\mu\varphi)}
&= \phi\,\frac{\partial \mathcal L_m}{\partial(\nabla_\mu\varphi)}
  + \phi_T\,\mathcal L_m\,\frac{\partial T}{\partial(\nabla_\mu\varphi)}
= \big(-\phi+2\phi_T\,\mathcal L_m\big)\,\nabla^\mu\varphi,\\
\frac{\partial (\phi\,\mathcal L_m)}{\partial \varphi}
&= \phi\,\frac{\partial \mathcal L_m}{\partial\varphi}
  + \phi_T\,\mathcal L_m\,\frac{\partial T}{\partial\varphi}
= -\big(\phi+4\phi_T\,\mathcal L_m\big)\,V_{,\varphi}.
\end{align*}
Euler--Lagrange then gives the exact equation
\begin{equation}
\nabla_\mu\!\left[\big(-\phi+2\phi_T\,\mathcal L_m\big)\,\nabla^\mu\varphi\right]
+\big(\phi+4\phi_T\,\mathcal L_m\big)\,V_{,\varphi}=0.
\label{eq:KG-exact}
\end{equation}
It is convenient to define
\begin{equation}
\mathcal D \equiv -\phi+2\phi_T\,\mathcal L_m,
\qquad
\mathcal S \equiv \phi+4\phi_T\,\mathcal L_m,
\end{equation}
so that \eqref{eq:KG-exact} becomes
\begin{equation}
\nabla_\mu(\mathcal D\,\nabla^\mu\varphi)+\mathcal S\,V_{,\varphi}=0.
\label{eq:KG-compact}
\end{equation}
In the GR/IR limit $\phi\to 1$, $\phi_T\to 0$, one finds $\mathcal D\to -1$ and $\mathcal S\to 1$, hence
\(-\Box\varphi+V_{,\varphi}=0\), i.e. \(\Box\varphi - V_{,\varphi}=0\), as expected.

\subsection*{Homogeneous FLRW form}
For a spatially homogeneous field $\varphi=\varphi(t)$ in FLRW,
\[
\mathcal L_m=-\tfrac12\dot\varphi^2 - V,\qquad
T=\dot\varphi^2-4V,
\]
and \eqref{eq:KG-compact} reduces to
\begin{equation}
\mathcal D\,(\ddot\varphi+3H\dot\varphi)
+\dot{\mathcal D}\,\dot\varphi
-\mathcal S\,V_{,\varphi}=0.
\label{eq:KG-FLRW}
\end{equation}
With the rational ansatz used in the main text,
\begin{equation}
\phi(T)=1+\frac{\beta\,T}{1+\gamma\,T},
\qquad
\phi_T(T)=\frac{\beta}{(1+\gamma T)^2},
\end{equation}
one substitutes $T=\dot\varphi^2-4V$ and $\mathcal L_m=-\tfrac12\dot\varphi^2 - V$ in $\mathcal D$ and $\mathcal S$.

\subsection*{Effective sources and bounce conditions}
Projecting the effective tensor on FLRW yields
\begin{equation}
H^2=\frac{\kappa}{3}\,\phi\,\rho,
\qquad
\dot H=-\frac{\kappa}{2}\,\big(\phi-2\phi_T\,\rho\big)\,\dot\varphi^2,
\qquad
\rho=\tfrac12\dot\varphi^2+V.
\end{equation}
A nonsingular bounce requires $H=0$ and $\dot H>0$. Since $\rho>0$, the bounce point satisfies
\begin{equation}
\phi(T_b)=0,\qquad 
T_b=\dot\varphi_b^{\,2}-4V(\varphi_b)=-\frac{1}{\beta+\gamma}\quad(\beta\neq 0,\ \beta+\gamma\neq 0),
\end{equation}
and smoothness follows if
\begin{equation}
\dot H\big|_b=\kappa\,\phi_T(T_b)\,\rho_b\,\frac{\dot\varphi_b^{\,2}}{2}>0
\quad\Longrightarrow\quad
\phi_T(T_b)=\frac{\beta}{(1+\gamma T_b)^2}>0\ \ \Rightarrow\ \ \beta>0.
\end{equation}
Thus the scalar-field realization reproduces the smooth, finite-density bounce found for the fluid case, and reduces to GR in the infrared.

\end{document}